**Imaging and Spectroscopy of Domains of the Cellular Membrane by Photothermal-Induced Resonance.**

Luca Quaroni

*Department of Physical Chemistry and Electrochemistry, Faculty of Chemistry, Jagiellonian University, 30-387, Kraków, Poland*

*Institute of Nuclear Physics, Polish Academy of Sciences, 31-342, Kraków, Poland*

**Abstract**

We use photothermal induced resonance (PTIR) imaging and spectroscopy, in resonant and non-resonant mode, to study the cytoplasmic membrane and surface of intact cells. Non-resonant PTIR images apparently provide rich details of the cell surface. However we show that non-resonant image contrast does not arise from the infrared absorption of surface molecules and is instead dominated by the mechanics of tip-sample contact. In contrast, spectra and images of the cellular surface can be selectively obtained by tuning the pulsing structure of the laser to restrict thermal wave penetration to the surface layer. Resonant PTIR images reveal surface structures and domains that range in size from about 20 nm to 1 µm and are associated to the cytoplasmic membrane and its proximity. Resonant PTIR spectra of the cell surface are comparable to far-field IR spectra and provide the first selective measurement of the IR absorption spectrum of the cellular membrane of an intact cell. In resonant PTIR images, signal intensity, and therefore contrast, can be ascribed to a variety of factors, including mechanical, thermodynamic and spectroscopic properties of the cellular surface. While PTIR images are difficult to interpret in terms of spectroscopic absorption, they are easy to collect and provide unique contrast mechanisms without any exogenous labelling. As such they provide a new paradigm in cellular imaging and membrane biology and can be used to address a range of critical questions, from the nature of membrane lipid domains to viral membrane fusion.

**Introduction**

Techniques for cellular and subcellular imaging have a prominent place in the analytical repertoire of biomedical scientists. Front place is taken by visible light microscopy, in all its incarnations from white light transmission microscopy to fluorescence microscopy. Resolution is generally limited by optical diffraction, although the development of super-resolved optical microscopy has extended it to the single molecule level. [1] While visible microscopy provides a wealth of morphological information under easily accessible and physiologically relevant conditions, molecular information generally requires exogenous labelling, which increases experimental complexity and limits applications.

The highest resolution is provided by electron microscopy, reaching a few nanometers for subcellular imaging.[2] However, the requirement for sample freezing and possible beam damage exclude measurements at environmental conditions, notably of live samples.

Scanning probe microscopy techniques, most notably variants of Atomic Force Microscopy (AFM), are seeing increased application to cellular studies, including live cells, with constantly improving performance.[3] They fill a specific and increasingly varied niche of the cell imaging landscape in providing unique access to a range of topographical, mechanical and electrical properties, occasionally with resolution down to the single macromolecule. Their main drawback is imposed by the requirement for contact or proximity between sample and probe, which requires indentation for intracellular studies, with a few exceptions.

In recent years, an increasing role has been occupied by spectroscopy-based microscopy techniques. Spectroscopic imaging relies on the local measurement of absorption, emission, or inelastic scattering of electromagnetic radiation by the sample to obtain spatially-resolved, molecular or atomic information. Local composition, orientation, dynamics, electronic, atomic and molecular properties, and intermolecular interactions can all be assessed, depending on the wavelength in use and the underlying physical process. All of them rely on endogenous spectroscopic properties of the sample, without the need of external labels, and are therefore suitable for the characterization of untreated and occasionally live samples.

Among spectromicroscopy techniques, IR absorption spectroscopy has seen extensive application to the study of biomolecules,[4] including complex biological matter, such as tissues and cells.[5] Despite years of intensive developments, two directions are still at open for development, the use of IR absorption to study living or functional samples[6] and the measurement of subcellular structures.[7] The latter capability clashes against the resolution limits imposed by diffraction on conventional optical measurements in the far field, of the order of the wavelength, approximately 2.5 – 25 µm in the mid infrared (mid-IR) spectral region commonly used for molecular studies. While such resolution allows investigation of single eukaryotic cells, and of their larger subcellular entities (the nucleus, vacuoles) it prevents application to most subcellular structures and to single prokaryotic cells. Several techniques have been developed or adapted to extend the resolution beyond the diffraction limit. An early development was the introduction of Scanning Near Field Optical Spectroscopy (SNOM). The technique was first demonstrated using visible light[8] and later extended to the mid-IR region.[9] It allows measuring samples as thin as single phospholipid bilayers with nanoscale resolution,[10] but the need to operate in a transmission configuration limits application to thicker samples, such as eukaryotic cells. A scattering SNOM configuration (s-SNOM) allows nanoscale IR spectroscopy by using the light scattered at the interface between a nanoscale metallic tip and the sample.[9] s-SNOM has been successfully used in the spectroscopy and imaging of viruses, single protein molecules,[11] single phospholipid bilayers, down to about 1000 phospholipid molecules,[12] and erythrocyte membranes.[13] The main constraint of the technique is its limitation to surface studies. Another technique, PiFM (Photo-induced Force Microscopy)[14] relies on measuring the force between induced optical dipoles at the interface between a metallized AFM tip and the sample.[14] PiFM allows both spectroscopic investigation of the sample and imaging with spatial resolution limited by tip size.[15] Similarly to s-SNOM, the technique is presently restricted to surface studies.

In contrast to other techniques, Photothermal - Induced Resonance (PTIR) spectroscopy (often termed AFM-IR) relies on the photothermal effect to induce local sample expansion following light absorption.[16] In PTIR the expansion is detected by a mechanical detection scheme instead of an optical one, using an AFM probe in contact with the sample. In the most common implementation, with pulsed excitation, the rapid deflection of the cantilever excites its resonant modes and the resulting oscillation is recorded as a voltage by the standard four-quadrant AFM detector.[17] PTIR is used as a spectroscopic technique, by scanning the wavelength of the light source while recording the deflection of the cantilever. The spectral response has been shown to track the IR absorption spectrum of the sample recorded with far-field optics. PTIR can be used as an imaging technique, by performing an AFM scan while simultaneously exciting the sample at one wavelength. An alternative imaging approach is the collection of spectral hypercubes, by recording extended PTIR spectra in a two-dimensional array of points and using the intensity of bands in each spectrum to reconstruct an image. PTIR is currently seeing increased applications in the study of biological samples with nanoscale resolution. Tissue, single cells, subcellular structures and purified biomolecules have all been investigated.[7,17–19,20]

An alternative scheme used to probe the photothermal response is to monitor the pressure and density gradients associated to the thermal expansion zone, as in Infrared Photoacoustic Microscopy (IR-PAM). The thermal expansion is detected by the deflection of an optical laser beam, yielding a spatial resolution that is limited by the wavelength of the probing beam, in the optical range down to UV wavelengths, instead of that of the exciting beam, in the infrared range.[21]

PTIR is the photothermal technique that provides the highest spatial resolution. We have recently shown that the spatial resolution of PTIR is limited by the interplay between the size of the photothermal expansion zone and the size of the tip.[22] By increasing frequency and duration of excitation pulses, it is possible to achieve conditions of super-resolution relative to the wavelength of the thermal wave and resolution values in the 10s of nanometers range become accessible.

The most challenging feature of PTIR arises from the indirect mechanical detection scheme of light absorption, which combines multiple factors in signal generation, in addition to the spectroscopic properties of the sample. The mechanics of tip-sample interaction [23–25] and the thermal properties of the sample and its environment [22] all contribute to signal intensity and even to spatial resolution.

Additional constraints are given by the limited sensitivity of non-resonant PTIR spectroscopy, which requires samples thicker than 100 nm, and the large depth response, which extends beyond 1 µm. Because of these constraints, successful application of non-resonant PTIR to the study of cellular samples has been reserved to the study of extended structures, such as lipid droplets and beads in the cell cytoplasm.[26] The limitations can be partly overcome by operation under resonant conditions, where the pulsing frequency of the laser matches a resonant frequency of the cantilever, giving enhancement proportional to the Q factor of the resonance. Resonant-mode PTIR extends the sensitivity to molecular monolayers,[27] but does not provide the selectivity required to separate the response of a membrane from that of the whole cell. However, recent results have shown that depth response, surface selectivity, sensitivity and resolution can be controlled by tuning the pulse structure of the excitation laser,[22] opening the way to the use of PTIR for the selective measurement of cell membranes. In the present work, we control the pulse structure of the excitation to selectively probe for the first time the cell surface and the cytoplasmic membrane, both in imaging and in spectromicroscopy experiments. We also discuss limitations and advantages of using the technique for spectroscopic and imaging studies of the cellular membrane.

# Experimental

*Cell Sample Preparation*

Epithelial cheek (buccal) cells were collected from the author by performing a cheek swab with a cotton bud, transferred to $CaF_2$ or ZnSe optical windows (25 mm diameter and 1 mm thickness; Crystran, Poole, UK) and measured without further treatment, as previously described.[19]

*PTIR Measurements*

PTIR measurements were carried out on a nanoIR2 instrument from Anasys Bruker (Santa Barbara, CA, USA). A Daylight Solutions MIRCat 1300 external cavity quantum cascade laser (QCL) and an optical parametric oscillator (OPO) laser were used for excitation, both in the top-down configuration. For all the experiments in this work, the beam is polarized vertically (electric field polarized 20° to the normal) and is focused to an elliptical spot approximately 100–200 μm long. Silicon probes with a 20 nm gold coating (PR-EX-nIR2-10, Anasys Bruker) and an overall tip diameter of 100 nm, were used for AFM and PTIR measurements.

The setup of the instrument is described in Figure 1. The basic design relies on multiple light sources, used for sample excitation, that are interfaced with an AFM head via an optical coupling box. In the configuration used in these experiments an OPO laser and a QCL are used. The coupling box contains optics to filter the beam, polarize it, measure its power, and steer it to the AFM head. The AFM is a modified conventional AFM design where the beam is focused on the sample in an upside-down geometry at an incident angle of approximately 70°.

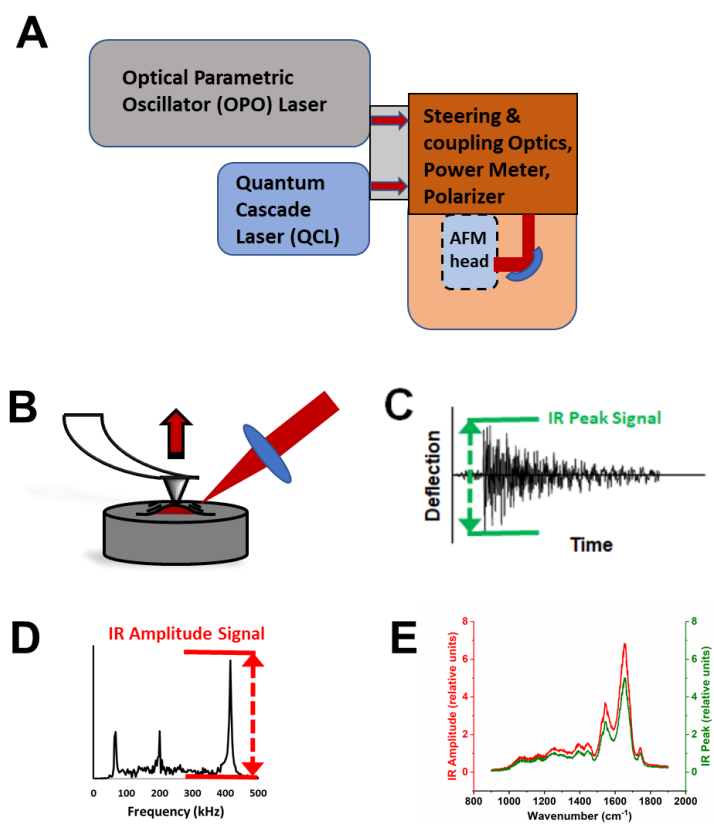

*Figure 1. PTIR experiment and setup. A. Scheme of instrumental setup. The instrument is based on a conventional AFM head integrated with laser light sources and a set of optics for coupling, shaping, and monitoring of the beam. B. The sample compartment. The sample is located at the focal point of the beam and under contact with the AFM probe. Illumination is with a top-down optical configuration. Sample irradiation at an absorbing wavelength causes sample expansion and is detected by the deflection and oscillation of the cantilever. C. In non-resonant operation, the PTIR signal can be quantified using the maximum peak-to-peak amplitude of the cantilever oscillation, as extracted from the ringdown of the cantilever. This is called the IR Peak output. D. In non-resonant operation, the PTIR signal can also be quantified using the amplitude of a resonance peak, among the resonances of the excited cantilever. This is called the IR Amplitude output. E. Comparison of the PTIR spectra obtained by plotting the IR Peak and the IR Amplitude signal as a function of wavenumber.*

*Non-resonant PTIR*

Non-resonant operation involves excitation of the sample at an absorbing wavelength with a series of short laser pulses to excite multiple cantilever resonances. The OPO laser used in the present experiments delivers 10 ns pulses at 1 kHz frequency. The pulsing frequency is lower than any of the resonant frequencies of the cantilever, which allows for the full decay of cantilever oscillations between pulses. Oscillations of the cantilever are recorded by the AFM detector as a decaying deflection signal and their amplitude of this oscillation is used to quantify the photothermal response of the sample. The decaying deflection signal can be processes along two different channels to provide a measure of the PTIR response. In the IR Peak channel, the peak-to-peak amplitude of the oscillation is plotted as a function of wavelength to obtain a PTIR spectrum for a stationary tip position (Figure 1C). Alternatively, the IR Peak channel can be plotted as a function of tip position at a fixed wavelength to obtain a PTIR map. In the IR Amplitude channel, the Fourier Transform of the IR Peak channel is used to obtain the full frequency spectrum of the oscillating resonant modes of the cantilever (Figure 1D). One resonance is then selected via the application of a gaussian filter and its amplitude is plotted as a function of wavelength, for spectra, or of tip position, for maps. IR Peak and IR Amplitude spectra are similar (Figure 1E), except for total amplitude and signal-to-noise levels. Non-resonant measurements were performed using the control and analysis package Analysis Studio version 2 from Anasys Bruker. For the present work PTIR spectra in non-resonant mode were collected by using the signal in the IR Amplitude channel and selecting the resonance at 64 kHz was selected for all spectral measurements using a gaussian filter with 30 kHz bandwidth. Laser power varying between 0.2 mW and 1.0 mW was used for excitation while the wavelength was scanned over the 900 cm$^{-1}$ - 3000 cm$^{-1}$ spectral range. Under these conditions, the temperature increase in the sample is estimated to be of the order of 1 K (see Supporting Information), which allows measurements for an extended time without significantly affecting the sample. Images in non-resonant PTIR operation were collected by scanning the AFM probe at a rate of 0.1 Hz or 0.08 Hz and co-averaging 16 or 8 scans respectively. The averaging is optimized to ensure that collection of the PTIR signal is equal to or faster than collection of the AFM signal. Images have pixels ranging in size from 0.8 nm to 15.6 nm, ensuring that AFM resolution is limited by tip size. The IR Peak channel and the IR Amplitude channel were used to plot PTIR maps. The resonance at 195 kHz was selected for all imaging measurements using a gaussian filter with 50 kHz bandwidth.

*Resonant PTIR*

Resonant operation provides greater signal intensity by matching the pulsing frequency of the laser to a specific cantilever resonance. A signal enhancement of the order of the Q factor of the cantilever is obtained. In addition, tuning of the pulsing frequency allows control of the depth response and resolution of the measurement. In the present setup, frequency matching is performed by locking the pulsing frequency of the QCL to the selected cantilever resonance. In this configuration, the PTIR signal is only extracted from the Amp 2 channel. The channel provides the amplitude of the selected resonance after selection by a gaussian filter with 50 kHz bandwidth, similarly to the IR Amplitude channel in non-resonant operation. Non-resonant measurements were performed using the control and analysis package Analysis Studio version 3.14 or later from Anasys Bruker.

The pulsing frequency, pulse length and pulse spacing were chosen to obtain the desired in-plane resolution and depth response, while matching a cantilever resonance, as previously specified.[22] The AFM probe was raster scanned at 0.4 Hz with a pixel resolution in the X and Y directions corresponding to a 6.25 nm or 7.8 nm pixel. 256 pulses were averaged for each pixel. Power at the sample was measured by the reference detector of the instrument and was 0.1–0.5 mW in the spectral region used for this investigation. PTIR images were not normalized to laser power. For resonant mode imaging, the PLL algorithm was used when specified in the text (Figure 4) to correct for changes in resonance frequency following changes in tip–sample contact and ensure that resonant conditions are maintained throughout the scan.

PTIR spectra in resonant mode were collected by using the same conditions for excitation as for imaging experiments. The tip was kept stationary at one location while the wavelength was scanned for the desired spectral range.

## Results and Discussion

*Non-Resonant PTIR Measurements*

Non-resonant PTIR was previously used to study intact primary epithelial buccal cells.[19] Epithelial buccal cells are a convenient system for methodological studies on cells because of their availability and ease of collection. They are frequently used in testing optical microscopes and for the demonstration of contrast and resolution performance in cellular imaging. Furthermore, the cells can be studied intact, without the need for fixation and in air, under which conditions they have shown to be remarkably stable to PTIR investigation. Preliminary PTIR measurements of buccal cells have been reported and have revealed an unusually strong response from unidentified spheroidal structures, presumably organelles or lipid droplets.[19] The present work extends the study of these cells using high resolution AFM and PTIR imaging, in attempt to investigate the cell membrane. Figure 2 shows the example of a buccal cell investigated by non-resonant PTIR. Supporting Figure 2 shows the cell mounted in the sample compartment. Figure 2A provides an overview of the cell topography in AFM contact mode. The nucleus is clearly visible in the middle of the cell and corresponds to the highest topographic location of the cell, similarly to adherent cells. Figure 2B shows a higher resolution scan of a small portion of the cell surface on top of the nucleus, indicated by the frame in panel A. The surface is highly irregular, corresponding to the folding of the cell membrane and associated polymers on top of the cell. The yellow arrows point at locations where PTIR (IR Amplitude) spectral measurements were subsequently performed. The resulting spectra are shown in Figure 2C, while Figure 2D shows the PTIR image at 2921 cm$^{-1}$ of the same area recorded synchronously with the AFM scan.

The spectra of Figure 2C show dominant bands in the Amide I and Amide II regions, around 1650 cm$^{-1}$ and 1545 cm$^{-1}$ respectively. Some of the spectra also show a weak but sharp absorption band between 1710 cm$^{-1}$ and 1750 cm$^{-1}$. Weaker bands are seen at lower frequencies. In the specific case of intact buccal cells, it has already been shown that non-resonant PTIR spectra qualitatively reproduce the band pattern of far-field FTIR spectra of fixed[7] and untreated[19] cells and of the most abundant cellular components. The spectra are also similar to non-resonant PTIR spectra recorded in some locations of fixed fibroblasts.[7]

The PTIR image of Figure 2D shows the response of the sample when exciting in an absorption band characteristic of long alkyl chain containing molecules, which includes acyl lipids and fatty acids. As already reported, the PTIR signal in the IR Peak output is not proportional to cell thickness and the highest intensity is given by unidentified organelles.[19] Their observation is confirmed by imaging with the IR Amplitude output. They appear as spots with higher intensity in several locations of the image and have been highlighted by the color scheme in Figure 2D. Comparison of the spectra in Figure 2C with the measurement location reveals that all spectra measured at the location of an organelle show the absorption band at 1710-1750 cm$^{-1}$, however this is not seen when the spectrum is measured away from the organelles themselves.

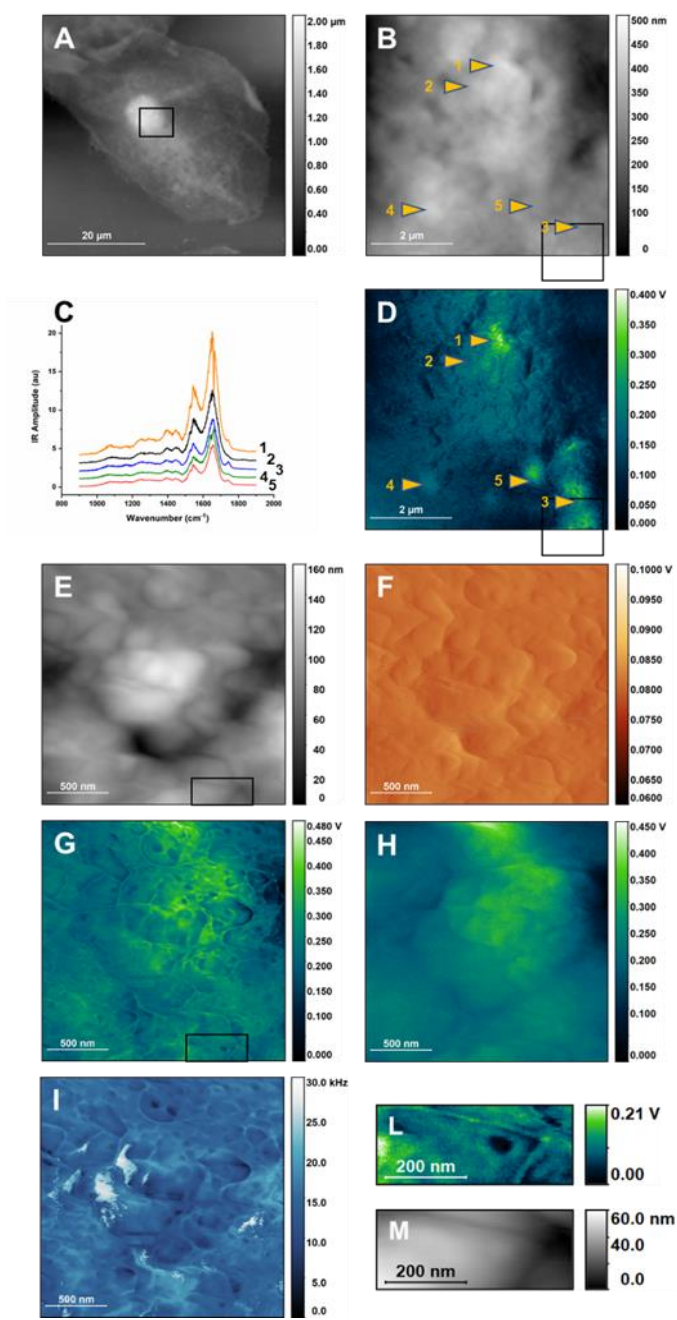

*Figure 2. Non-resonant PTIR images of buccal cells and of their cellular membrane. A. AFM Height image of a buccal cell. The frame shows the region mapped in B and D B. Higher resolution AFM Height image of a portion of the cell shown in A. Arrows point at locations where PTIR spectra of panel C were measured. The frame shows the region mapped at higher resolution in panels E to I. C. Non-resonant PTIR spectra of selected locations in panels B and D. The frame shows the region mapped at higher resolution in panels E to I. D. Non-resonant PTIR (IR Amplitude) map at 2921 cm$^{-1}$ of the same region as in B. The frame shows the region mapped at higher resolution in panels E to I. Arrows point at locations where PTIR spectra of panel C were measured. E. Higher resolution AFM Height image of a region of B. The frame indicates the region that is graphically expanded in M. F. AFM Deflection image of the same*

*region as in E. I. Non-resonant PTIR (IR Amplitude) image at 2921 cm$^{-1}$ of the same region as in E. The frame indicates the region that is graphically expanded in M. H. Non-resonant PTIR (IR Peak) image at 2921 cm$^{-1}$ of the same region as in E. I. Contact frequency image of the same region as in E. L. Cutout of the PTIR (IR Amplitude) image at 2921 cm$^{-1}$ in I, lower right corner. M. Cutout of the AFM Height image in E, lower right corner. Panels A and B have been adapted from reference $^{19}$, in accordance with the terms and conditions of the Creative Commons Attribution (CC BY-NC-ND 4.0) license. Part of the experimental dataset used for this figure was also used in reference $^{19}$ for a different type of analysis.*

Overlaid to the organelles is a fine pattern with rich details, which shows much higher contrast when measuring on top of the organelles themselves. The pattern is not observed using the IR Peak output and has not been reported before. Contrast decreases rapidly away from the high intensity location, indicating that it is related to the intensity of the signal itself. To further investigate the fine pattern, higher resolution AFM and PTIR images were collected on top of one of the organelles. Figures 2E and 2F show the resulting AFM Height and Deflection images. Figures 2G and 2H compare the PTIR images recorded from the IR Amplitude and IR Peak channels, respectively. Figure 2I shows the change in the cantilever contact frequency when scanning through the sample. Details extracted from Figure 2C and Figure 2E are shown in the panels of Figure 2L and Figure 2M respectively.

The PTIR image from the higher intensity IR Amplitude channel (Figure 2G) reveals with high contrast several structures. They include apparent filaments, globular structures and clusters of the latter that form a crown pattern (One of the latter is detailed in Figure 2L). The lowest apparent size of these structures is as small as 10-20 nm, observed for the filaments. This pattern is absent in the corresponding AFM height and deflection maps (Figure 2E and Figure 2F), showing that it is not directly related to changes in topography. However, the IR Amplitude map does accurately track the variation in contact resonant frequency during the PTIR scan (Figure 2I). Changes in the contact frequency arise from changes the mechanical properties of tip-sample contact, which appear to modulate the contrast observed in IR Amplitude maps. The map in Figure 2H shows the corresponding PTIR image recorded using the IR Peak channel, corresponding to the peak-to-peak difference in the unfiltered ringdown trace. The IR Peak signal shows no high-resolution patterns in the image and tracks instead the AFM height and deflection images. Comparison of Figure 2 panels confirms that contrast in Figure 2G and Figure 2I (and also the finer details in Figure 2D) is dominated by mechanical factors, namely changes in the contact frequency of the AFM probe during scanning. This is particularly obvious when analyzing the structures that look like filaments. Comparison with the Height maps indicates that their location corresponds to that of the crevices formed by folding of the surface. We conclude that the "filaments" are not specific cellular structures but represent instead the change in contact resonance frequency that accompanies the sharp change in tip-sample contact as the tip crosses over a gap. The use of resonance filtering to isolate single frequencies in the IR Amplitude channel implies that frequency shifts from mechanical factors bring the resonance to the edge of the filter bandwidth and lead to a decreased output signal. Therefore, frequency shifts are manifest as intensity changes in the IR Amplitude map but not in the IR Peak map.

The crown like pattern in Figure 2L consists of a ring of five or six hemispherical structures around a central core. The pattern cannot be observed in the matching AFM height panel of Figure 2M, where it appears as an undifferentiated convex structure. Therefore, PTIR contrast must arise from differences in mechanical properties of the various regions. Several of these patterns can be observed in Figure 2G or Figure 2I. Further work is necessary to identify these structures. They could be large protein complexes or clusters of structured membrane domains with very different mechanical properties. In the general case, higher PTIR signal is expected from stiffer structures, indicating that the central core is softer than the surrounding crown.

The high spatial resolution of the patterns in Figure 2E and Figure 2H confirms that the origin of the contrast is not in the spectroscopic properties of the membrane itself. The in-plane resolution and depth response of PTIR depends on the pulse structure of the excitation laser, including the pulsing frequency. Like all measurements based on the photothermal effect, the limiting spatial resolution depends on the volume of the photothermal expansion zone, and is different from resolution in AFM images, which is limited by tip size. High spatial resolution in PTIR, of the order of 10's nm, and shallow depth response can be obtained only when using pulsing frequencies higher than 100 kHz and pulse duration longer than 100 ns, to restrict the size of the photothermal expansion volume.[22] In contrast, measurements in Figure 2 are recorded using a laser pulsing frequency of 1 kHz, with 10 ns pulses. In-plane spatial resolution at this pulsing frequency is estimated at 160 nm (Supporting Information) and the depth response is expected to cover the whole thickness of the cell.[28] A photothermal measurement would not allow resolving the fine structures shown in Figure 2G and Figure 2L. Their observation can be accounted for only if contrast has a mechanical origin and resolution is comparable to tip size. While Figure 2G and Figure 2L, and the finer details of Figure 2D, may provide a fascinating representation of the cellular surface, they do not represent IR absorption, but changes in the mechanical properties of the tip-sample contact. As the mechanical properties of the surface are also affected by structures below the immediate surface layer, it is possible that some of the objects in Figure 2G and Figure 2L are not in direct contact with the tip but are deeper within the cell.

Careful observation of Figure 2D allows observing the relationship between the mechanical contrast of fine surface images and overall signal intensity. Contrast is high in the more strongly absorbing regions of the sample and decreases away from them. We conclude that IR absorption from the interior of the cell is responsible for the PTIR response, and the signal is further modulated by changes in the mechanical properties of the sample, which provide the finer contrast details.

Figure 3 compares the response in the different channels by showing a profile extracted from the same position in the images from different channels. The profile is selected to run across convex and concave regions of the sample (Figure 3A), including gaps, with a vertical excursion of approximately 20 nm (Figure 3E, black trace). The profile runs in a region of relatively low deflection (Figure 3B), corresponding to a few mV, and appreciable changes in deflection can be measured only at the crossing of a gap (Figure 3E, orange trace). Along the same profile, the IR Peak (Figure 3C) and IR Amplitude (Figure 3D) images show much higher signal. The large difference between the deflection signal and the IR Peak and IR Amplitude signals confirms that there is little or no crosstalk between PTIR and deflection channels, despite some similarities between the corresponding images. The IR Amplitude profile (Figure 3E, red trace) has similar intensity as the IR Peak profile (Figure 3E, green trace) but shows the modulation caused by the changing contact frequency, as previously discussed. The IR Amplitude profile is also less noisy than the IR Peak and Deflection profiles, in part because the signal is obtained by filtering a single resonance, thus removing the noise from outlying frequency regions. It is notable that the IR Amplitude profile is substantially different from the Height profile. Local maxima in IR Amplitude signal correspond to gaps in the surface topography, while some local minima correspond to apexes in topography. The discrepancy leads further support to the proposition that the IR Amplitude signal is heavily affected by the modulation of tip-sample contact.

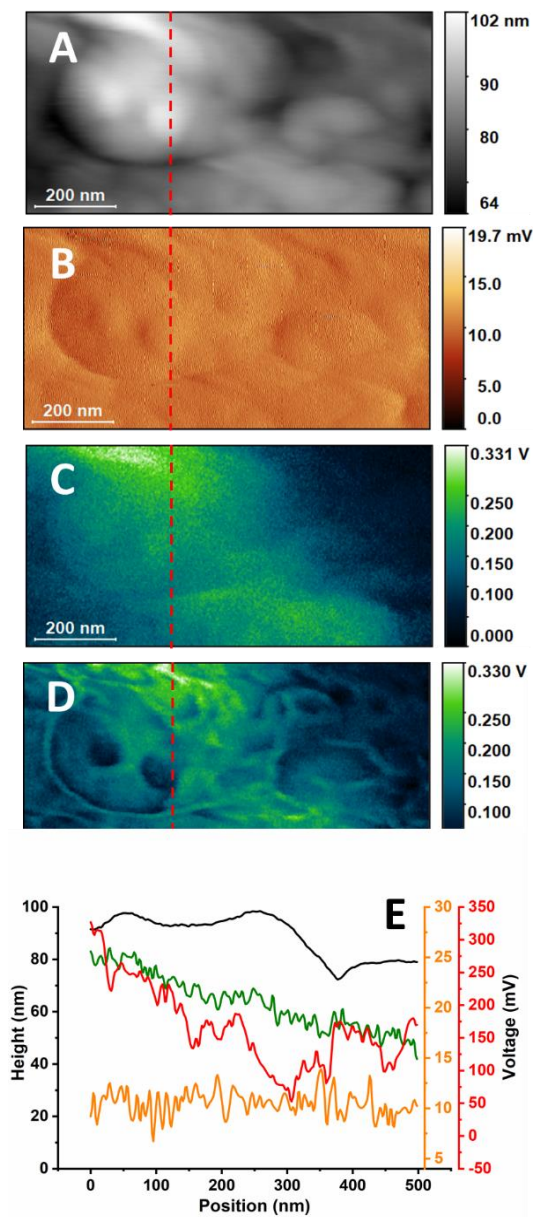

*Figure 3. Line profiles in different non-resonant images of the cell. The figure compares line profiles extracted from the same location (red dashed line) of images recorded synchronously on different channels using non-resonant excitation. A. Height image. B. Deflection image. C.*

*IR Peak image. D. IR Amplitude image. E. Comparison of profiles. Height, black trace, left black scale. Deflection, orange trace, right orange scale. IR Peak, green trace, right red scale. IR Amplitude, red trace, right red scale.*

*Resonant PTIR Measurements*

In its simplest description, a cellular membrane comprises a phospholipid bilayer, about 5-6 nm thick, and associated biopolymers, including proteins and polysaccharides, leading to a thickness of the order of 10-20 nm. The conditions used for the non-resonant PTIR experiments in Figure 2 do not provide the surface selectivity nor the sensitivity to measure the cellular membrane itself or its immediate proximity. To achieve the required selectivity and sensitivity, measurements were performed at higher laser pulsing frequency and under resonant conditions, matching the pulsing frequency of the laser to a cantilever resonance. Resonant operation increases the signal by the Q factor of the specific resonance and extends sensitivity to ultrathin samples, down to a monomolecular layer,[27] while the use of higher pulsing frequencies provides increased surface selectivity by reducing the depth extension of the photothermal expansion volume.[22]

Figure 4 shows the result of this approach to the imaging of the surface of a buccal cell. PTIR imaging at lower frequency provides images with very low structure and higher noise levels, while imaging at higher frequency reveals a wealth of detail and lower noise images. At the highest frequency used in Figure 4E and Figure 4F (821 kHz) the photothermal depth response extends to about 100 nm (Supporting Information), according to the classical description of thermal wave propagation.[29] The non-linear thermal response observed from soft matter,[23] including cellular samples,[22,30] can further reduce this value by up to one order of magnitude, bringing it to 10 - 20 nm, roughly the thickness of the cell membrane and its associated biopolymers.

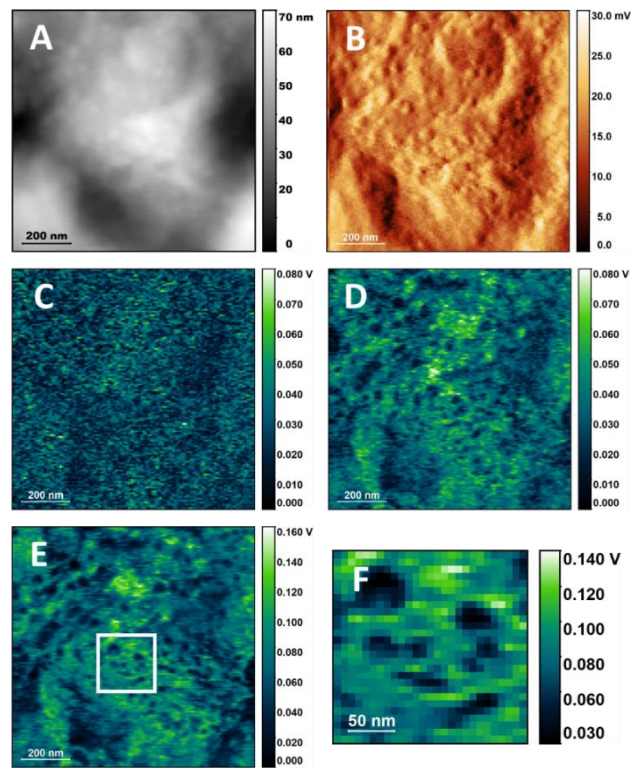

*Figure 4. PTIR images of the cellular surface of buccal cells with resonant-mode excitation. A. AFM height image of a cell surface. B. AFM deflection image of the same region as in A. C – E. PTIR images of the same region as in A using 1654 $cm^{-1}$ excitation and 65 kHz, 312 kHz and 821 kHz pulsing frequency respectively. F. Detail of E showing surface structures. No resonance tracking was implemented for these images.*

Figures 4D to 4F show a wealth of structures, which can be classified roughly into two groups, particles and extended domains. Particles are circular or elongated structures that range in size from about 20 nm to 70 nm (3 to 8 pixels of 7.8 nm each), as seen in the detail of Figure 4E. Most of the particles are smaller than tip size (50 nm radius) and close to the value reported for the limiting in-plane resolution of the measurement under the present conditions.[22] Many of them can be associated to convex protuberances in the AFM height map of Figure 4A. Based on their dimension and shape, they could be tentatively identified with larger complexes of proteins or other biopolymers, smaller lipid phases within the membrane or protein lipid assemblies. Residual deposits from the biopolymers in saliva cannot be ruled out. Interestingly, the size of the smaller structures corresponds to that of the smaller membrane domains reported from mobility experiments based on optical trapping,[31] in agreement with predictions from the updated fluid mosaic model [32] and the lipid rafts model.[33]

Extended domains are larger portions of the cell surface that are characterized by uniform PTIR intensity, such as the clear region that occupies the lower half of Figure 4E (or the whole of Figure 4F), about 1 µm in size. They can be differentiated by the higher intensity of the signal relative to the surroundings, suggesting greater stiffness. The description suggests a tentative identification with the extended domains denominated coalesced lipid rafts.[34] In cellular membranes, lipid rafts can be directly imaged using fluorescence microscopy of tagged molecules.[35,36] To date few reports describe the optical imaging of lipid domains in intact cells without the use of exogenous labels, and rely on inelastic scattering techniques, such as Raman microscopy,[37] and Tip Enhanced Raman Spectroscopy (TERS) imaging.[38] The present work demonstrates the capability of label-free visualization of phase separation in the membrane of intact cells using PTIR imaging.

While control of the depth of the photothermal expansion allows us to avoid signal from the bulk of the cell and obtain selective signal form the cellular surface, it does not remove the dependence of the PTIR signal from the mechanical properties of the sample. Images in Figure 4 were collected without actively matching the pulsing frequency and contact resonance frequency, which also introduces an additional mechanical contribution to contrast. Nonetheless, even when using correction tools, such as the PLL algorithm of the instrument, mechanical contributions are presently unavoidable because of the very design of the PTIR experiment, which relies on tip-sample contact. Efforts are underway to correct it post-measurement [39] or to reduce it by operating in tapping mode.[40] Such developments may eventually allow the separation of spectroscopic and mechanical components in the PTIR signal of some sample typologies. However, improvements in our understanding of how the mechanical properties of deep-seated structures affect surface measurements will be necessary before such corrections can be extended to more complex heterogeneous samples, including eukaryotic cells.

Spectroscopic and mechanical properties of the sample are not the only contributions to the contrast observed in Figure 4. It has been proposed that the thermal properties and geometry of both sample and environment can affect not only signal but also contrast and resolution, by affecting the rate of thermal relaxation.[22,41] In the case of cellular samples, crevices and concave regions give rise to stronger PTIR signal than apexes and convex regions. This is observed for non-resonant (Figure 2 and Figure 3) and for resonant excitation (Figure 4), where weaker signal is provided by the smaller convex structures.

The resulting description of PTIR signal generation reveals a complexity that arises from the micromechanical detection scheme, which introduces a dependence of the signal from the mechanical properties of the sample and tip, while the measurement of the photothermal effect introduces a dependence from the thermomechanical and optical properties of the sample. These contributions to the signal are difficult to disentangle when measuring structurally and compositionally complex samples, such as cells. For these samples and contrary to what is often assumed, contrast in a PTIR image with single wavelength excitation cannot be simply reduced to differences in light absorption and care must be exercised when using the images for quantitative compositional analysis.

When the IR absorption of the sample is of interest, single location PTIR spectra, rather than images, are a more reliable reporter of the spectroscopic properties of the sample. During such measurement, the tip remains stationary in one location while the wavelength of the excitation source is scanned. Provided that all factors that affect signal intensity are wavelength independent, PTIR spectra track the corresponding far-field absorption spectra. While the former condition needs to be verified case by case, a similarity between PTIR and far-field spectra is often observed for many samples, including cells.[42] ~~In the specific case of intact buccal cells, it has already been shown that single spectra qualitatively reproduce the band pattern of far-field FTIR spectra of fixed and untreated cells and of the most abundant cellular components.~~

We now expand the use of single resonant PTIR spectra to perform spectroscopic studies of the surface of the cell. Spectra of the cellular membrane have been recorded by using higher pulsing frequency and longer pulses, to maximize the selectivity of the response relative to the cell interior, as in imaging experiments. Spectra of the 1500 – 1800 cm$^{-1}$ spectral region were measured for different locations and are shown in Figure 5, together with AFM height (Figure 5 A) and PTIR images at 1655 cm$^{-1}$ (Figure 5 B) where the measurement positions have been marked. The overlapping pattern of negative sharp bands due to atmospheric water vapor absorption in the optical cavity of the QCL has not been removed by post processing to avoid spectral distortion.

The spectra in Figure 5C have similar or identical appearance and the same bands are observed in all of them, with similar relative intensity. The main difference is the overall intensity of the signal, with the intensity of the strongest peak at 1655 cm$^{-1}$ changing by about 100 mV among spectra. The PTIR map at 1655 cm$^{-1}$ (Figure 5B) also shows changes in intensity of the order of 100 mV, demonstrating that the overall intensity of the spectra tracks the intensity of the signal throughout the map. This is further confirmation that contrast in single wavelength PTIR maps is affected by factors other than spectroscopic absorption.

To further test the relationship between PTIR images and spectra of the cell surface, Figure 5D and Figure 5E compare PTIR images recorded at two different wavelengths, 1654 cm$^{-1}$ and 1622 cm$^{-1}$. The images show different levels of signal intensity and contrast but are very similar in all other aspects, and the same structures can be observed in both. The difference in signal intensity qualitatively tracks the difference in IR absorption of the sample at the two wavenumbers, in agreement with the spectra of Figure 5C. The comparison allows us to reaffirm that PTIR signal generation does increase with sample absorption, despite the contribution of other factors.

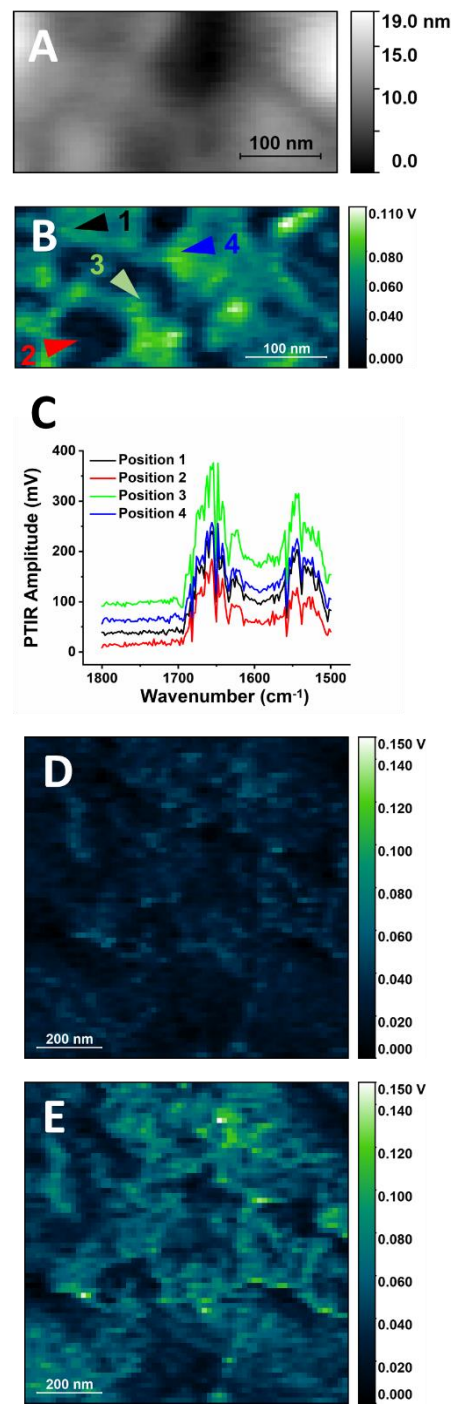

*Figure 5. Resonant-mode PTIR spectra of the cellular membrane of a buccal cell. A. AFM Height image of a cellular surface region. B. Resonant-mode 820 kHz PTIR image at 1655 cm$^{-1}$ of the same region. The arrows point to the position used for spectral measurements. C. Resonant-mode PTIR spectra recorded in the marked locations. The vertical scale of the spectra has been staggered to facilitate comparison. D. Resonant-mode 820 kHz PTIR image at 1622 cm$^{-1}$ of a region of the cell surface. E. Resonant-mode 820 kHz PTIR image at 1654 cm$^{-1}$ of the same region as in D.*

The spectra in Figure 5C are qualitatively similar to the ones observed in far-field IR absorption spectra of cells, with two dominant peaks around 1545 cm$^{-1}$ and 1655 cm$^{-1}$.[6] The two peaks are typically assigned to the Amide I and Amide II modes of the amide group of polypeptide chains, although some amino acid side chains, lipids and other molecules can also contribute.[43] One difference from cellular far-field spectra, and from protein spectra in general, is the relative intensity of the two peaks. This is comparable in Figure 5C, while in far-field IR spectra the Amide I peak is markedly more intense than the Amide II peak. This discrepancy has already been noted in fixed cells [7] and its origin is presently unclear. The relative intensity of Amide I and Amide II bands in non-resonant PTIR spectra of these same cells (Figure 2C) is similar to the one observed in far-field measurements. The difference may arise from the widely different volumes measured in the two experiments. Non-resonant measurements have a deeper response from the low pulsing frequency, of the order of the micrometer [28] and probe the whole cell thickness. Resonant measurements are performed at high pulsing frequency and probe only the surface layer of the cell. According to this interpretation, the difference between Figure 2C and Figure 5C could arise from the presence of surface species that absorb at 1545 cm$^{-1}$ or from anisotropy in the orientation of absorbing molecules in the surface layer. However, it cannot be presently ruled out that the difference could also arise from differences in the signal generation mechanism between resonant and non-resonant experiments.

Another notable difference between from non-resonant spectra is the absence of absorption bands in the 1700 – 1750 cm$^{-1}$ range of resonant spectra. The carbonyl groups of protonated acids and their esters, including fatty acids, phospholipids and triglycerides, absorb here. Their absorption can be appreciated in the IR transmission spectra of lipid rich samples, including cells, but also in the PTIR spectra of large subcellular structures[19] and the remains of cellular organelles after fixation.[5] An absorption band at 1742 cm$^{-1}$ is observed in non-resonant PTIR measurements (Figure 2C) and is associated to the still unidentified organelles that generate a strong PTIR signal in these cells. The band is also observed in the non-resonant PTIR spectra of organisms that accumulate acyl-lipids.[26] In contrast, the ester carbonyl band has not been observed in PTIR spectra of supported purple membranes[18] and is very weak in sSNOM spectra of single bilayers of pure phospholipids,[12] while measurements of protein Amide bands in single membranes have proven easy, both by PTIR[18] and sSNOM.[44] Therefore the failure to observe the ester band in the spectra of Figure 5 is fully consistent with the relative weakness of the absorption when compared to Amide bands in a single membrane sample. The observation confirms that the spectra of Figure 5C arise from the cellular membrane and its proximity.

The results support the viability of single PTIR spectra for studies of the cell surface, in contrast to the difficulties encountered in imaging applications. We demonstrate it by using cells that are untreated, to ensure the integrity of the cellular membrane. The measurements can also be extended to fixed and adherent cells, provided that care is taken in assessing the spectroscopic contribution from any associated chemical modification (e.g. lipid removal during fixation, or the presence of poly-lysine as an adhesive). In general, it is expected that these conclusions are not limited to the case under investigation but can be extended to other complex samples.

While the cells used in the present experiments were not tested for viability, live epithelial cheek cells can be collected with the same protocol and successfully cultured. By proper control of environmental conditions, such as humidity, it is foreseeable that live epithelial cheek cells could be measured by PTIR in the very same conditions described in this work, without resorting to an aqueous environment.[19] The mild conditions of the measurement, corresponding to photothermal heating of the order of 1 K, ensure sample viability for extended periods of time. Furthermore the same method could be extended to other epithelial cells that can be maintained under similar conditions. Developments in this direction would make PTIR a valuable technique for biochemical studies of this cell class. Relevant bioanalytical applications can also be envisioned with relation to pathologies of various epithelial tissues, including the oropharyngeal and nasopharyngeal cavity, the pulmonary alveolar surface and eye membranes.

**Conclusions**

We show that non-resonant PTIR measurements of intact epithelial eukaryotic cells provide images and spectra of the core of the cell. Surface associated structures are also observed in non-resonant PTIR images, but their contrast is defined not by their IR absorption but by shifts in the contact frequency of the cantilever, which provides a representation of the mechanical properties of the tip-sample contact. Selective imaging and spectroscopy of the surface of the cell is possible by operating in resonant-mode at higher laser pulsing frequency, thus limiting the photothermal depth response to a surface layer. Even under these resonant conditions, PTIR image contrast is still determined by a combination of mechanical, thermodynamic and spectroscopic properties of the sample. Surface geometry also contributes to signal intensity, with concave surfaces providing stronger signal than convex surfaces. Unfortunately, non-spectroscopic contributions to signal intensity are often neglected in the interpretation of PTIR imaging data, at the cost of unreliable conclusions. Care must be exercised when using PTIR images for compositional analysis of the sample, by wrongly assuming that intensity is a simple direct representation of light absorption.

In terms of spectroscopic analysis, operation at high pulsing frequency provides spectra of the surface components of the cell that can be qualitatively related to far-field IR absorption spectra. Strong signals are observed from bands in the Amide I and Amide II position, tentatively assigned to molecules in the membrane or in its proximity. However, despite the shallow depth response used in the measurement, no signal is observed from the carbonyl absorption of lipids and fatty acids, indicating that the contribution of such molecules is much weaker.

The complexity of the PTIR response is a limitation when the aim is the spectroscopic analysis of the sample, but it can be turned into an advantage for imaging purposes. The variety of mechanisms involved in contrast generation results in rich images, where individual structures can be differentiated based on the combined effect of their spectroscopic, mechanical and thermal properties. In the present experiments, we use it to image components of the cellular surface and identify structures with properties expected for membrane domains, as described in the fluid mosaic [32] and lipid raft models.[33]

Overall, PTIR imaging delivers the resolution of electron microscopy, the combined contrast of scanning probe and spectroscopic microscopy, and the sampling ease of light microscopy, making it widely applicable to the biomedical field, for fundamental research and diagnostic purposes. One currently relevant example is the potential application to the study of the mechanisms of internalization of bacterial and viral pathogens in epithelial host cells. The sampling technique used for the present measurements bears strong resemblance to the one used in testing for COVID related infection of the oropharyngeal cavity. PTIR imaging measurements could be easily adapted to the rapid diagnostic screening of infected samples.

**Acknowledgments**

L.Q. was supported by the European Union's Horizon 2020 research and innovation programme under the Marie Skłodowska-Curie grant agreement no. 665778, managed by the National Science Center Poland under POLONEZ contract 2016/21/P/ST4/01321, and by an OPUS16 grant from the National Science Center Poland under contract 2018/31/B/NZ1/01345.

The research was performed using equipment purchased in the frame of the project cofunded by the Małopolska Regional Operational Programme Measure 5.1 Krakow Metropolitan Area as an important hub of the European Research Area for 2007–2013, project no. MRPO.05.01.00-12-013/15. The author thanks prof. dr. hab. Czesława Paluszkiewicz, prof. dr. hab. Wojciech Kwiatek, dr. Katarzyna Pogoda, dr. Natalia Piergies, and dr. Ewa Pięta of IFJ-PAN for nanoIR2 instrumentation access, support and discussion. The author is also grateful to prof. dr. hab. Maria Nowakowska for support throughout the Polonez project and to Dr. Miriam Unger, Dr. Anirban Roy, Dr. Qichi Hu and the staff of Anasys/Bruker for information and discussion about the instrumentation.

**References**

1    L. Möckl, D. C. Lamb and C. Bräuchle, *Angew. Chemie Int. Ed.*, 2014, **53**, 13972–13977.

2    A. J. Koster and J. Klumperman, *Nat. Rev. Mol. Cell Biol.*, 2003, **4**, 6–10.

3    A. Berquand, C. Roduit, S. Kasas, A. Holloschi, L. Ponce and M. Hafner, *Micros. Today*, 2010, **18**, 8–14.

4    V. A. Lorenz-Fonfria, *Chem. Rev.*, 2020, **120**, 3466–3576.

5    L. Quaroni, *Infrared Phys. Technol.*, 2018, 102779.

6    L. Quaroni and T. Zlateva, *Analyst*, 2011, **136**, 3219.

7    L. Quaroni, K. Pogoda, J. Wiltowska-Zuber and W. M. Kwiatek, *RSC Adv.*, 2018, **8**, 2786–2794.

8    U. Dürig, D. W. Pohl and F. Rohner, *J. Appl. Phys.*, 1986, **59**, 3318–3327.

9    B. Knoll and F. Keilmann, *Nature*, 1999, **399**, 7–10.

10  A. M. Siddiquee, I. Y. Hasan, S. Wei, D. Langley, E. Balaur, C. Liu, J. Lin, B. Abbey, A. Mechler and S. Kou, *Biomed. Opt. Express*, 2019, **10**, 6569.

11  I. Amenabar, S. Poly, M. Goikoetxea, W. Nuansing, P. Lasch and R. Hillenbrand, *Nat. Commun.*, 2017, **8**, 14402.

12  A. Cernescu, M. Szuwarzyński, U. Kwolek, P. Wydro, M. Kepczynski, S. Zapotoczny, M. Nowakowska and L. Quaroni, *Anal. Chem.*, 2018, **90**, 10179–10186.

13  A. Blat, J. Dybas, M. Kaczmarska, K. Chrabaszcz, K. Bulat, R. B. Kostogrys, A. Cernescu, K. Malek and K. M. Marzec, *Anal. Chem.*, 2019, acs.analchem.9b01536.

14  I. Rajapaksa, K. Uenal and H. K. Wickramasinghe, *Appl. Phys. Lett.*, 2010, **97**, 3–5.

15  J. Jahng, D. A. Fishman, S. Park, D. B. Nowak, W. A. Morrison, H. K. Wickramasinghe and E. O. Potma, *Acc. Chem. Res.*, 2015, **48**, 2671–2679.

16  A. Dazzi, R. Prazeres, F. Glotin and J. M. Ortega, *Opt. Lett.*, 2005, **30**, 2388.

17  A. Dazzi and C. B. Prater, *Chem. Rev.*, 2017, **117**, 5146–5173.

18  V. Giliberti, M. Badioli, A. Nucara, P. Calvani, E. Ritter, L. Puskar, E. F. Aziz, P. Hegemann, U. Schade and M. Ortolani, 2017, **1**, 1–9.

19  L. Quaroni, *Molecules*, 2019, **24**, 4504.

20  F. S. Ruggeri, G. Longo, S. Faggiano, E. Lipiec, A. Pastore and G. Dietler, *Nat. Commun.*, 2015, **6**, 7831.

21  J. Shi, T. T. W. Wong, Y. He, L. Li, R. Zhang, C. S. Yung, J. Hwang, K. Maslov and L. V. Wang, *Nat. Photonics*, 2019, **13**, 609–615.

22  L. Quaroni, *Anal. Chem.*, 2020, **92**, 3544–3554.

23  A. Dazzi, F. Glotin and R. Carminati, *J. Appl. Phys.*, , DOI:10.1063/1.3429214.

24  D. E. Barlow, J. C. Biffinger, A. L. Cockrell-Zugell, M. Lo, K. Kjoller, D. Cook, W. K. Lee, P. E. Pehrsson, W. J. Crookes-Goodson, C. S. Hung, L. J. Nadeau and J. N. Russell, *Analyst*, 2016, **141**, 4848–4854.

25  S. Mittal, K. Yeh, L. Suzanne Leslie, S. Kenkel, A. Kajdacsy-Balla and R. Bhargava, *Proc. Natl. Acad. Sci. U. S. A.*, 2018, **115**, E5651–E5660.

26  A. Deniset-Besseau, C. B. Prater, M. J. Virolle and A. Dazzi, *J. Phys. Chem. Lett.*, 2014, **5**, 654–658.

27  F. Lu and M. A. Belkin, *Opt. Express*, 2011, **19**, 19942–7.

28  B. Lahiri, G. Holland and A. Centrone, *Small*, 2013, **9**, 439–445.

29  A. Rosencwaig, *Science*, 1982, **218**, 223–228.

30  L. Wang, C. Zhang and L. V. Wang, *Phys. Rev. Lett.*, 2014, **113**, 1–5.

31  C. Eggeling, C. Ringemann, R. Medda, G. Schwarzmann, K. Sandhoff, S. Polyakova, V. N. Belov, B. Hein, C. Von Middendorff, A. Schönle and S. W. Hell, *Nature*, 2009, **457**, 1159–1162.

32  G. L. Nicolson, *Biochim. Biophys. Acta - Biomembr.*, 2014, **1838**, 1451–1466.

33  K. Simons and E. Ikonen, *Nature*, 1997, **387**, 569–572.

34  K. Simons and M. J. Gerl, *Nat. Rev. Mol. Cell Biol.*, 2010, **11**, 688–699.

35  S. Miersch and B. Mutus, *Curr. Anal. Chem.*, 2006, **3**, 81–92.

36  A. K. Kenworthy, B. J. Nichols, C. L. Remmert, G. M. Hendrix, M. Kumar, J. Zimmerberg and J. Lippincott-Schwartz, *J. Cell Biol.*, 2004, **165**, 735–746.

37  M. Pilarczyk, L. Mateuszuk, A. Rygula, M. Kepczynski, S. Chlopicki, M. Baranska and A. Kaczor, *PLoS One*, , DOI:10.1371/journal.pone.0106065.

38  M. Richter, M. Hedegaard, T. Deckert-Gaudig, P. Lampen and V. Deckert, *Small*, 2011, **7**, 209–214.

39  S. Kenkel, A. Mittal, S. Mittal and R. Bhargava, *Anal. Chem.*, 2018, **90**, 8845–8855.

40  J. Mathurin, A. Deniset-Besseau and A. Dazzi, *Acta Phys. Pol. A*, 2020, **137**, 29–32.

41  A. N. Morozovska, E. A. Eliseev, N. Borodinov, O. S. Ovchinnikova, N. V. Morozovsky and S. V. Kalinin, *Appl. Phys. Lett.*, 2018, **112**, 033105.

42  J. Mathurin, E. Pancani, A. Deniset-Besseau, K. Kjoller, C. B. Prater, R. Gref and A. Dazzi, *Analyst*, 2018, **143**, 5940–5949.

43  L. Quaroni, I. Benmessaoud, B. Vileno, E. Horváth and L. Forró, *Molecules*, 2020, **25**, 336.

44  I. Amenabar, S. Poly, W. Nuansing, E. H. Hubrich, A. A. Govyadinov, F. Huth, R. Krutokhvostov, L. Zhang, M. Knez, J. Heberle, A. M. Bittner and R. Hillenbrand, *Nat. Commun.*, 2013, **4**, 1–9.